# Can we predict mixed grain boundaries from their tilt and twist components?

## Author information

Wei Wan [1, 2], Changxin Tang [2, *] and Eric R. Homer [3]

[1] *Institute for Advanced Study, Nanchang University, Nanchang, 330031, China*

[2] *Institute of Photovoltaics, Nanchang University, Nanchang, 330031, China*

[3] *Department of Mechanical Engineering, Brigham Young University, Provo, 84602, USA*

[*] *Corresponding author, Email address:* tcx@ncu.edu.cn

## Abstract

One of the major challenges towards understanding and further utilizing the properties and functional behaviors of grain boundaries (GB) is the complexity of general GBs with mixed tilt and twist characters. Here, we report the correlations between mixed GBs and their tilt and twist components in terms of structure, energy and stress field by computationally examining 7040 silicon GBs. Such correlations indicate that low angle mixed GBs are formed through the reconstruction mechanisms between their superposed tilt and twist components, which are revealed as the energetically favorable dissociation, motion and reaction of dislocations and stacking faults. In addition, various complex disconnection network structures are discovered near the conventional twin and structural unit GBs, implying the role of disconnection superposition in forming high angle mixed GBs. By unveiling the energetic correlation, an extended Read-Shockley model that predicts the general trends of GB energy is proposed and confirmed in various GB structures across different lattices. Finally, this work is validated in comparison with experimental observations and first-principles calculations.

## 1. Introduction

Grain boundaries (GBs) separating individual crystals exhibit unique impacts on the structural and functional performances of crystalline materials, such as strength, plasticity, toughness, corrosion resistance and electronic activity [1–6]. Remarkable increments in these performances have been made by controlling the population of desired GB types, which emerged as a field called grain boundary engineering (GBE) [7, 8]. Demands of GBE for silicon materials have increased due to the rapid growth of the semiconductor and photovoltaic industries [9, 10]. Unfortunately, most of our knowledge about GBs is focused on FCC metals with low stacking fault energies [11–15]. To enable GBE for silicon materials, the structure-property relationships of various silicon GBs must be determined. In other words, we need to understand the basic GB structures and energetic properties that determine various GB behaviors (e.g., migration, diffusion, solute segregation and defect sink) [16–26].

Structures and properties of a given GB are jointly defined by the macroscopic and microscopic degrees of freedom (DOFs). The five macroscopic DOFs are known as the GB character, three of which define the misorientation axis between two crystals and the other two describe the boundary plane normal. For each unique macroscopic structural descriptor, numerous microscopic DOFs on the GB atomic arrangements can form a multiplicity of meta-stable structures, and their properties play a critical role in the material designs [27–29]. For accurately tracing GB structures at the atomic scale, both simulations and experimental methods are mutually complementary because some complex GB structures are very difficult for experimental access, and experimental methods would help to confirm simulation prediction [30–32]. Moreover, the growing use of machine-learning enables the characterization of complex interactions between GB structures and properties, while accelerating experimental and computational design and discoveries [33–39].

Historically in the studies about GBs in silicon [40–45] and other materials [46, 47], the five DOF GB characters were often investigated in simplified geometries like symmetric tilt or twist or even down to one independent DOF known as misorientation angle. The knowledge about the simple tilt/twist GBs and their properties is extensive, including the Read-Shockley relationship [48] that predicts the GB structures and energies as a function of misorientation angle, the Frank-Bilby equation (FBE) [49, 50] predicting dislocation structures of low angle grain boundaries (LAGBs) [51–53], the structural/polyhedral unit models [54–56] characterizing high angle grain boundary (HAGB) structures, the role of disconnection in shaping GB structures [57, 58], the particularity (e.g., high occurrence of frequency, representativeness of entire GB population) of low Σ (reciprocal density) Coincidence Site Lattice (CSL) GBs [12, 59], and the universality of GB structures among different FCC metals [60]. Beyond these findings, the topological analysis of the symmetry of 5D GB space yields a unique strategy named Fundamental Zone, which reveals the role of the boundary plane normal and the misorientation axis [61–64]. The latest computational approach [65] is capable of examining nearly the entire 5D GB space due to the rapid development of computer resources.

Although these reviewed studies almost constructed today's understandings of GB, they still lack comprehensive coverage or generalized approaches of the possible GB characters because an arbitrary GB is not limited to the widely studied symmetric tilt or twist types. Once geometrically available, the co-existence of symmetric tilt and twist DOFs, known as the mixed tilt-twist GB character, appears. Earlier works [66–70] have suggested an analytical method for studying this GB type through its decomposition into tilt and twist components with the nearest crystallographic distances, and addressing the correlations within. For example, low angle symmetric tilt grain boundaries (LASTGBs) are often considered dislocation arrays that fall in the prediction of FBE [71, 72], and low angle twist grain boundaries (LATwGBs) are reported as dislocation networks with quadrate, hexagonal or more complex topology [73–77]. Going further, it has been experimentally and numerically shown that the low angle mixed grain boundaries (LAMGBs) contain dislocation characteristics of their two components and involve complex structural reconstructions [66, 78–81]. Understanding such mechanisms is essential for the design and application of self-assembled nano-structures, which use the LAMGB as the nano-pattern template [78]. Meanwhile, although Wolf had contributed preliminary interpolations of mixed GB energies [67] using the tilt and twist decompositions, further knowledge about the mixed GB character and its structure-property relationships is still quite limited [82]. Therefore, some intriguing questions could be raised: 1) What is their common structural feature? 2) How does a property, such as energy or mobility, vary with the two DOFs? 3) Most importantly, can we transfer the knowledge about the one-DOF GBs to them?

In this atomistic study, we compute silicon mixed GB structures near the common (001), (011), (111) and (112) boundary planes with molecular mechanics. The unique protocol of this study is that the mixed GBs are studied by decomposing into the tilt and twist components, which allows us to understand the structural and energetic correlations among tilt, twist and mixed tilt-twist GB geometry. Contrary to previous works that typically addressed simple symmetric tilt and twist GBs, we analyzed a considerable amount of complex mixed GBs with two independent DOFs and established a generalized model on the basis of the classical Read-Shockley framework to characterize the mixed GB energies.

## 2. Methodology

### 2.1. Grain boundary modelling

Figure 1 shows the manner to characterize the GBs and their geometry. To identify a mixed GB character, we use three parameters $u$, $v$ and $w$ and three orthogonal orientations (misorientation axis $i$, tilt axis $j$ and boundary plane normal $k$) to set the rotation matrices of two crystals to $R_1 = [i_1\ j_1\ k_1]$ and $R_2 = [i_2\ j_2\ k_2]$, where $i_1 = ui + vj + wk$, $j_1 = -vi + uj$, $k_1 = -uwi - vwj + (u^2 + v^2)k$, $i_2 = ui - vj - wk$, $j_2 = vi + uj$ and $k_2 = uwi - vwj + (u^2 + v^2)k$. The mixed GB decomposes into a symmetric tilt grain boundary (STGB) with tilt angle $\theta = 2 \times \arctan(w/u)$ and a twist grain boundary (TwGB) with twist angle $\phi = 2 \times \arctan(v/u)$ at $ik$ and $ij$ planes, respectively. By varying the values of $w/u$ and $v/u$ from 0 to ∞, a 2D space, called mixed character space, is created, which symmetry is shown in Figure 1.

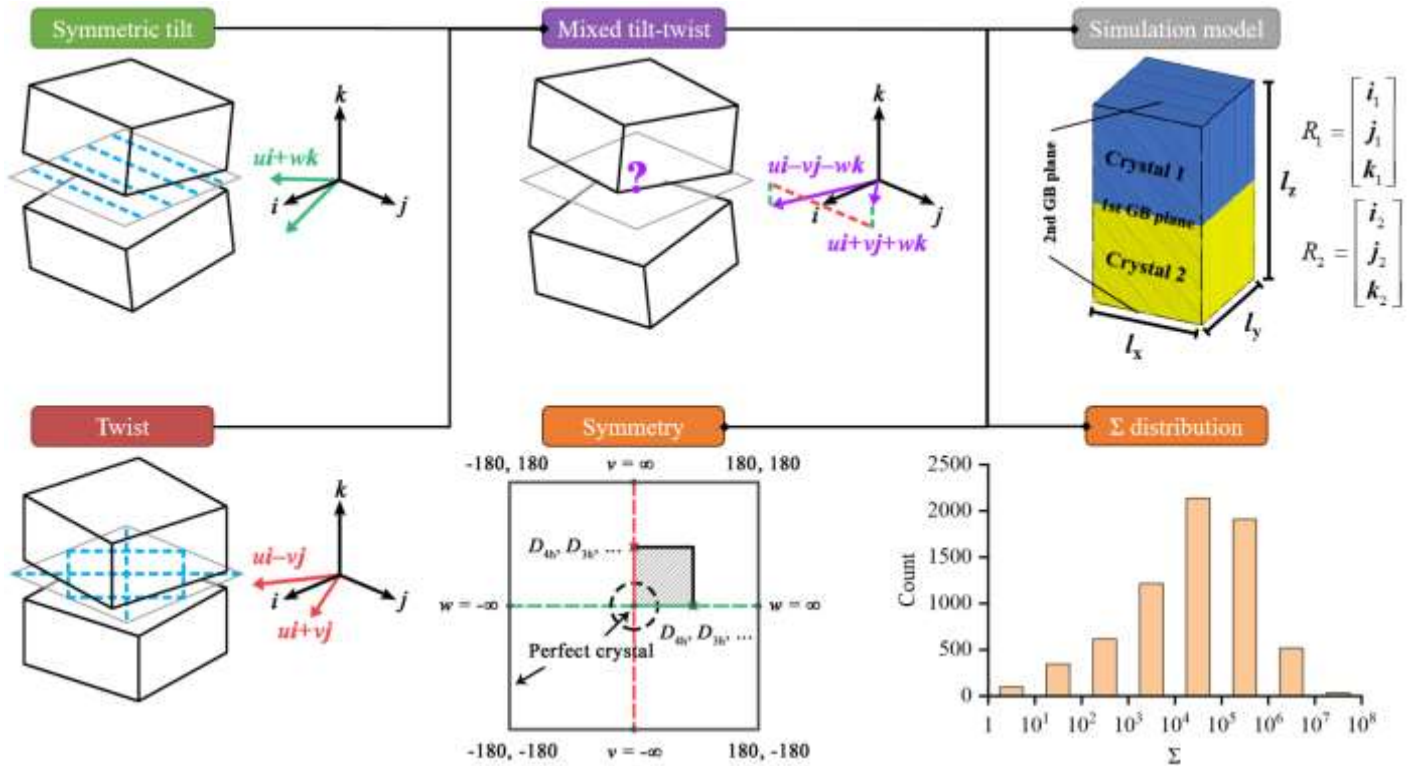

**Figure 1.** Geometry settings of the mixed tilt-twist GB character. The domain of the mixed character space is [-180°, 180°] × [-180°, 180°]. The domain center (0°, 0°) and four boundaries are perfect crystals or asymmetric tilt grain boundaries (depending on the selection of $k$ and $i$). LAMGB are found near the domain center (black dashed circle) and other positions

where the perfect crystals appear. STGBs and TwGBs are located in the 1D subsets [-180°, 180°] × [0°] and [0°] × [-180°, 180°], respectively. The symmetry of the axes $i$ and $k$ determines the perfect crystals and symmetric equivalence in the two mentioned 1D subsets (marked by the × symbol; For example, 90° for $D_{4h}$ symmetry and 180° for $D_{2h}$ symmetry). Note that different selections of $k$ and $i$ may yield the same GB character, and thus the mixed character space is unable to uniquely index a GB character. Σ values of this GB dataset range from 1 to $10^8$, quite high compared with a similar approach [65].

The mixed character space and the associated GB dataset are identified by their unique orientations $k$ and $i$ before applying geometry transformation of the tilt and twist combination. For example, a set of (001)/[100] mixed GB would use $k$ = (001) and $i$ = [100]. In this work. four combinations of $k$ and $i$ are considered and result in four mixed character spaces and GB datasets, including (001)/[100], (011)/[100], (111)/[110] and (112)/[110]. To form mixed GB characters, 32 (001)/[100] STGB are crossed with 32 (001) TwGBs, resulting in 1024 (001)/[100] GBs (i.e., ($w/u$, 0) crossed with (0, $v/u$) in the space). Similarly, 48 STGBs and 48 TwGBs are crossed to obtain 2304 (011)/[100] GBs, 48 STGBs and 32 TwGBs are crossed to obtain 1536 (111)/[110] GBs, 32 STGBs and 48 TwGBs are crossed to obtain 1536 (112)/[110] GBs. In all cases, the crossing provides systematic coverage of the mixed character space at ~ 3° × 3° resolution, and the generated GB characters are mostly mixed type. The lists of specific sample tilt and twist GBs are given in Supplementary Tables S1 to S8. Since the tilt and twist combination generates a diverse set of GB types, we define all of the different types and sub-types of GBs that are involved in the discussion in Table 1

Table 1. Classification of symmetric tilt, twist and mixed GBs in the results.

| Type | Full name | Sub-type | Full name & Description |
|---|---|---|---|
| STGB | Symmetric Tilt Grain Boundary | LASTGB | Low Angle Symmetric Tilt Grain Boundary comprised of dislocation array |
|  |  | HASTGB | High Angle Symmetric Tilt Grain Boundary comprised of structural unit or extended structure* |
|  |  | DAT | Disconnection Array on the coherent Twin boundary (no observation of incoherent type in this dataset) |
| ATGB | Asymmetric Tilt Grain Boundary | —— | an asymmetric tilt angle formed between the boundary plane and the symmetric plane |
| TwGB | Twist Grain Boundary | LATwGB | Low Angle Twist Grain Boundary comprised of dislocation network |
|  |  | HATwGB | High Angle Twist Grain Boundary comprised of structural unit or extended structure |
| Mixed GB | Mixed tilt-twist Grain Boundary | LAMGB | Low Angle Mixed Grain Boundary comprised of dislocation network |
|  |  | HAMGB | High Angle Mixed Grain Boundary comprised of structural unit or extended structure |
|  |  | LAMTGB | Low Angle Mixed Twin Grain Boundary comprised of twin grain boundary and disconnection network |

* While GBs are generally regarded as extended defects, we use the term "extended structure" to refer to more diffuse defects in atomic structures of GBs similar to the conception that the extended kite is a diffuse kite-like structure [27, 37].

LAMMPS [83] simulations are used to generate GB structures at zero temperature and pressure in a periodic box following the published sampling method [14, 65]. In total, 35.2 million structures are examined for 7040 GBs (6400 mixed, 320 symmetric tilt and 320 twist GBs). As illustrated in Figure 1, most of the GB Σ are above $10^3$ and the average Σ is approximately $4.4 \times 10^5$. To the authors' knowledge, these numbers, as well as the GB size and structural richness that they represent have shown an order of magnitude greater than most of the reviewed minimum GB energy datasets [11, 14, 28, 30, 32–35, 43–45, 60].

Only classical interatomic potentials could minimize so many GB structures of this size within acceptable costs. The authors adopted a modified Tersoff potential [84], which not only reproduces the elastic constants and generalized stacking fault energies of silicon but is also proven capable of modelling complex atomic bond environments and dislocation structures [66]. First-principles calculations [85–88] are used for parallel comparisons with atomistic simulations. The details are given in the Supplementary Information.

The dislocation analysis tool (DXA) implemented in Ovito [89, 90] is used to identify dislocation structures, setting the trial Burgers circuit length to 9 atom-to-atom steps and a default circuit stretchability.

GB energy $E_{GB}$ from atomistic simulation and first-principles calculation are defined as the following:

$$E_{GB} = \frac{\sum_i^N (E_i - E_{Coh})}{A_{GB}} \tag{1}$$

Where $N$ is the atom count of a GB in the simulation box (usually half of the box atoms because a box contains two GBs). $E_i$ is the energy of atom $i$, $E_{Coh} = -4.63$ eV is the cohesive energy of silicon atoms [91] and $A_{GB}$ is the GB area. The Virial stress tensor of each atom is computed to analyze stress fields.

## 2.2. First-principles validation

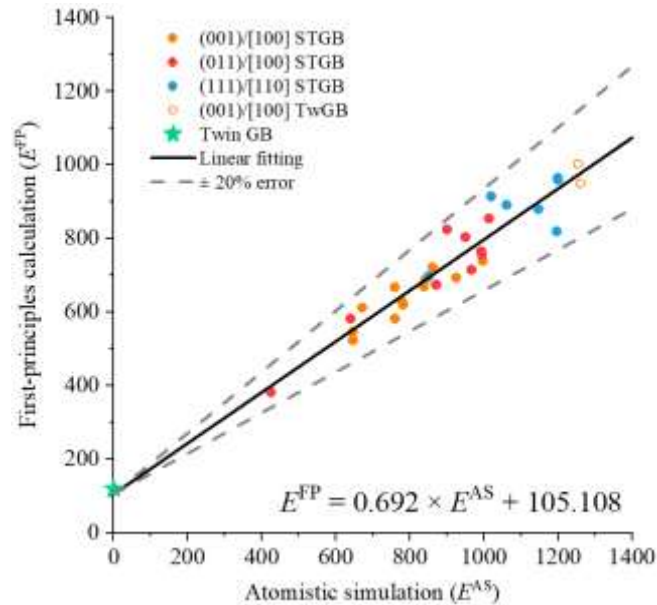

**Figure 2.** Comparison of GB energies between atomistic simulation and first-principles calculation. Energy from atomistic simulation $E^{AS}$ and energy from first-principles calculation $E^{FP}$ follows the linear relationship $E^{FP} = 0.692 \times E^{AS} + 105.108$.

To validate the GB energies presented here, low Σ GBs with simulation cell sizes that fall in the acceptable computational range are calculated using first-principles. The comparison of energies by first-principles and molecular statics is plotted in Figure 2. The simulated energy is in good linear scaling with the first-principles calculation (falls in ± 20% error lines), and suggests that the computed energy surface is reliable. However, it is noted that the twin GB energy in the atomistic simulation is nearly zero (0.04 mJ/m$^2$), while the first principles calculation gives an energy of 119.55 mJ/m$^2$. These results are consistent with the published data of silicon and carbon [45]. The energies would only need to be corrected with a linear relationship for comparison with first principles calculation. Also, this fact indicates that similar trends of the energy surface presented below will be observed if more accurate (but consuming) computational methods are deployed.

## 3. Results & Discussions

### 3.1. Grain boundary energy

Figures 3a to 3d show the (001)/[100], (011)/[100], (111)/[110] and (112)/[110] mixed GB energies as a continuous function of tilt angle $\theta$ and twist angle $\phi$, respectively. Inspection of GB structures, described in section 3.2, indicates that all GB energy surfaces can be divided into three parts: (i) LAMGBs comprised of dislocation network structures; (ii) LAMTGBs comprised of disconnection network structures; (iii) HAMGBs comprised of extended structures, which regions are divided by the black dashed line.

Nine unique LAMGB zones are identified by the Roman numerals I to IX and the three unique LAMTGB zones are identified by α, β and γ in Figures 3a to 3d. These zones are of principal interest because their nature allows the observation of how mixed GB structures are reconstructed from the tilt and twist GBs, as described in section 3.2. The corresponding energy trends are steadily, described in the extended framework of the Read-Shockley relationship in section 3.3. However, of additional interest here are special HAMGBs near the low energy STGBs or TwGBs.

The twelve zones of interest have elliptic shapes, some of which only show a quarter or a half of the ellipse due to the symmetry of the space. Specific types of LASTGB, LATwGB and disconnection-array-twin (DAT) structures (all of which are described in Table 1) form in these regions are identified in Figures 3a to 3d. For example, the DAT structures and LASTGBs are respectively observed passing through the twin GB and perfect crystal, and they represent the parts of the common 1D symmetric tilt or twist subsets that are frequently reported [11, 13, 14, 30, 32, 40–45, 73–75]. These elliptical zones all share a common feature of zero or near zero energy at the ellipse origin that increases slowest along constant $\theta$ or $\phi$ directions. The elliptic long and short axes that define the ten LAMGB and LAMTGB zones are determined by the core radius of dislocations in these LASTGBs, LATwGBs and DAT structures (i.e., dislocation core radius determines the curvature of each elliptic zone), where a detailed explanation is given in the work of Wan and Tang [66]. The energies of these structures are then influenced by the degree that STGB and TwGB structures reconstruct to form unique GB structures, as discussed in the following section.

The HAMGB zones that cover the majority of the energy surface have several notable features. First, there are band-like landscapes that cross horizontally and vertically. In some cases, these bands are remarkable due to the high energy values (marked as the high energy zones). In other cases, these bands form low energy cusps. These bands are also non-uniform, accompanied by smaller bands that form diagonally. In Figure 3a, the band connecting zones II and III even forms an arc. The energies of most HAMGBs are very close to the average energy of all examined GBs at around 1340 mJ/m$^2$, a quite high value because most simple silicon HASTGB energies are near 800 mJ/m$^2$.

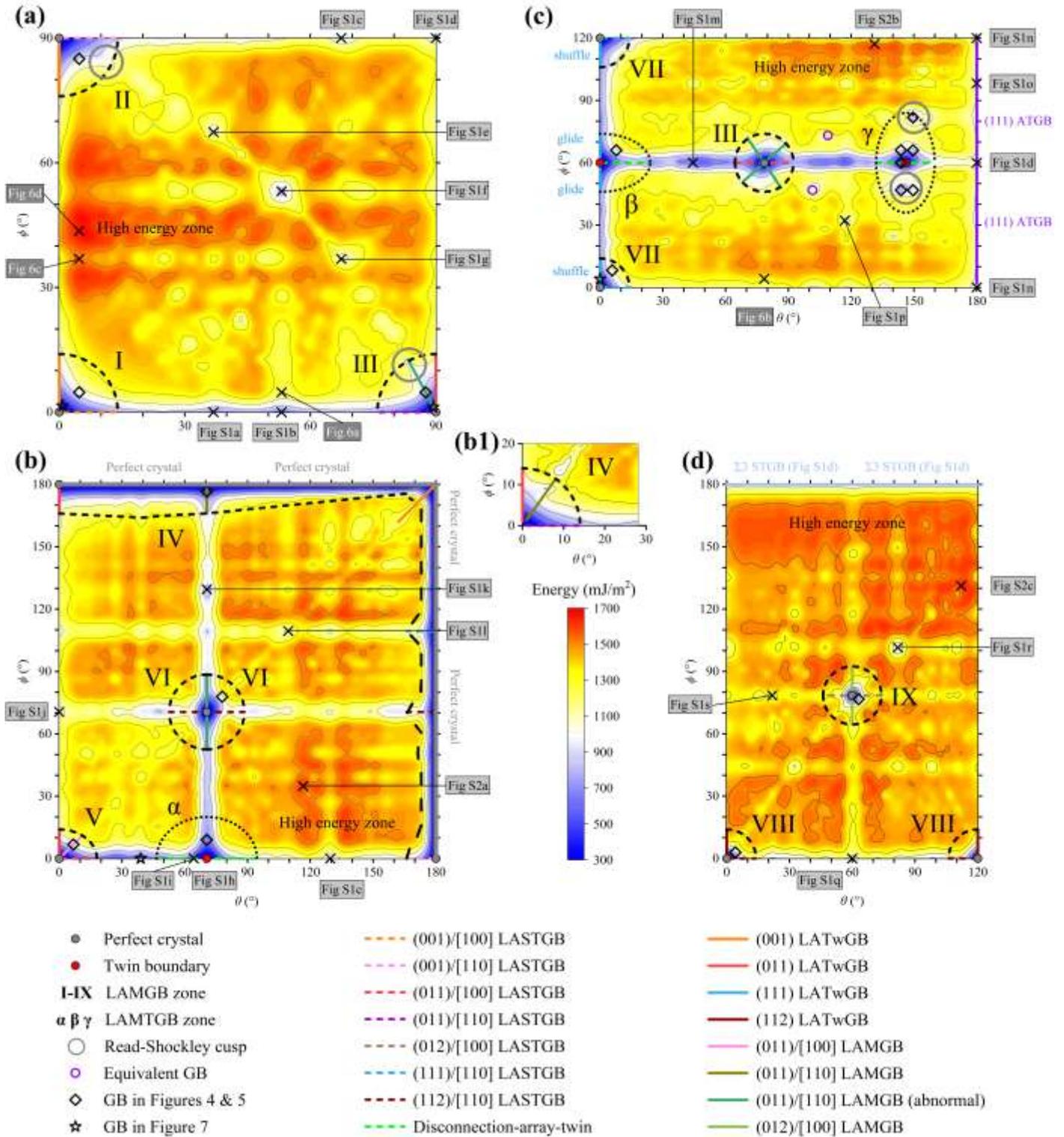

**Figure 3.** Energy surfaces of the examined mixed GBs as the functions of $\theta$ and $\phi$. (a) (001)/[100] mixed GB; (b) (011)/[100] mixed GB; (c) (111)/[110] mixed GB; (d) (112)/[110] mixed GB; (b1) A portion of (011)/[110] mixed GB. The mixed character space is non-Euclidean space with unknown curvature, and thus zone IV of (011)/[100] mixed GB in (b) is the expansion of zone IV of (011)/[110] mixed GB in (b1). Even in the LAMGB zones where the dislocation density is proportional to the tilt and twist angles, one should notice that the energy trends are not completely smooth and show Read-Shockley cusps.

In summary, the energy surfaces plotted in Figure 3 have characteristic features around specific structures, such as LAGB, twin GB, and perfect crystals, and generally have high energy in the HAMGB zone. Clearly there are cusps in the HAMGB zone that could have a better understanding in the future work. Additional observations and more geometry details for each energy surface are presented in Supplementary Information.

## 3.2. Grain boundary structure

In this section, we analyze how tilt and twist GB structures result in mixed GB structures for three mixed GB types: LAMGB, LAMTGB and special HAMGB near the low-energy STGBs or TwGBs. Experimental comparisons are also given for validation.

### 3.2.1. LAMGB & LAMTGB structures

Dislocation structures of nine LAMGBs in zones I to IX are plotted in Figure 4, where all LASTGBs and LATwGBs are dislocation arrays and dislocation networks (that may contain stacking faults), respectively. It is worth noting that the dislocation network topology (count array as a special network, similar to Figure 4a1) should not change once the ratio between the tilt and twist angles (tilt-twist ratio, $TTR$, defined as $|\theta/\phi|$ here) is determined, and the dislocation network size is scaled by misorientation angle (e.g., absolute values of $\theta$ and $\phi$). Both come from the inherent geometry definition of LAGBs (transferable among materials), and are described with more details in references [66, 75–80]. Formation mechanisms of the nine LAMGBs (Figures 4a2 to 4i2) can be considered as the superposition of their tilt (Figures 4a1 to 4i1) and twist (Figures 4a3 to 4i3) components after energetically favorable reconstructions (dislocation dissociation, motion and reaction). These results support other observations of superposition and reconstruction in some previously reported silicon LAMGBs [66]. The types and Burgers vectors of each dislocation segment are marked to make the reconstructions clearly traceable. For (001) LAMGBs shown in Figures 4a2 and 4b2, the reconstructions are based on ½<110> screw dislocations, which suffer segmentation of ½<110> mixed dislocation, and then glide half of their spacing. There are more complex reconstructions involving the stacking fault from (011) LATwGB in Figures 4c2, 4d2 and 4e2. ½<110> edge dislocations separate both stacking fault (equivalent to a <100> dislocation) and ½<110> screw dislocation of (011) LATwGB in two different orientations, which finally yields two different LAMGBs in Figures 4d2 and 4e2. Although the reconstruction mechanism governing (011) LAMGB structures is complicated and involves stacking fault, it follows the same with the (001) LAMGB once the stacking fault is approximated as a dislocation. Even for the complex LAMGBs in Figures 4g2, 4h2 and 4i2, the generation of ⅙<411>, ¼<111> and ⅓<221> mixed dislocations are also caused by the dissociation and reaction associated with ½<110> screw dislocation.

It should be noted that two LAMGB zones are considered special or abnormal. The first is zone III, where all LAMGBs do not contain infinite straight dislocation lines. The dislocation dissociation and reaction are the same with the LAMGBs in zone IV, but the ½<110> mixed dislocation moves to the edge of the stacking fault area and forms a zigzag pattern. Although zone III decomposes exactly into the same tilt and twist components as zone IV, it cannot be indexed by examining the (011)/[110] energy surface in Figure 3b. Instead, it can be indexed by examining the (001)/[100] energy surface, which suggests that other GB DOFs are involved. The second is zone VI because its decomposed twist component is LAMGB in zone III rather than LATwGB. LAMGBs in zone VI is the superposition of (112)/[110] LASTGB and (011)/[110] LAMGB where the LAMGB plays the role of twist component.

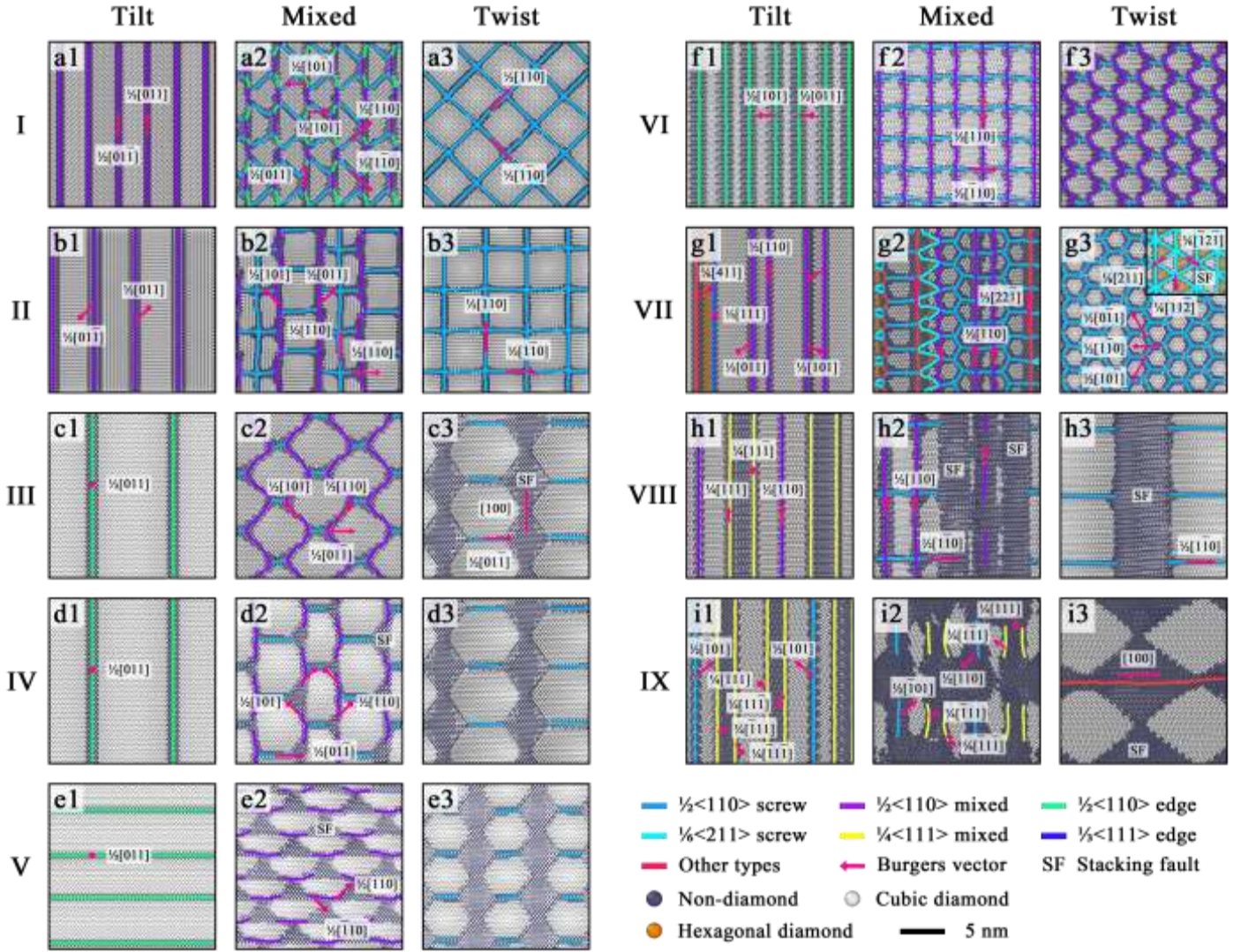

**Figure 4.** Dislocation structures of low angle symmetric tilt, twist and mixed GBs in the LAMGB zones I to IX of Figure 3. (a1) $\theta = 4.77°$ (001)/[100] LASTGB; (a2) $\theta = 4.77°$, $\phi = 4.77°$ (001)/[100] LAMGB; (a3) $\phi = 4.77°$ (001) LATwGB; (b1) $\theta = 3.38°$ (001)/[110] LASTGB; (b2) $\theta = 3.38°$, $\phi = 4.77°$ (001)/[110] LAMGB; (b3) $\phi = 4.77°$ (001) LATwGB; (c1) $\theta = 2.43°$ (011)/[110] LASTGB; (c2) $\theta = 2.43°$, $\phi = 3.39°$ (011)/[110] abnormal LAMGB; (c3) $\phi = 3.39°$ (011) LATwGB; (d1) $\theta = 2.43°$ (011)/[110] LASTGB; (d2) $\theta = 2.43°$, $\phi = 3.39°$ (011)/[110] LAMGB; (d3) $\phi = 3.39°$ (011) LATwGB; (e1) $\theta = 4.50°$ (011)/[100] LASTGB; (e2) $\theta = 4.50°$, $\phi = 4.50°$ (011)/[100] LAMGB; (e3) $\phi = 4.50°$ (011) LATwGB; (f1) $\theta = 7.91°$ (112)/[110] LASTGB; (f2) $\theta = 7.91°$, $\phi = 6.38°$ (112)/[110] LAMGB; (f3) $\theta = 4.51°$, $\phi = 6.38°$ (011)/[110] abnormal LAMGB (mixed GB as the twist component); (g1) $\theta = 5.84°$ (011)/[100] LASTGB; (g2) $\theta = 5.84°$, $\phi = 8.26°$ (011)/[100] LAMGB; (g3) $\phi = 8.26°$ (011) shuffle LATwGB (the glide variant is given in sub-figure); (h1) $\theta = 4.13°$ (112)/[110] LASTGB; (h2) $\theta = 4.13°$, $\phi = 2.92°$ (112)/[110] LAMGB; (h3) $\phi = 2.92°$ (112) LATwGB; (i1) $\theta = 3.05°$ (012)/[100] LASTGB; (i2) $\theta = 3.05°$, $\phi = 1.77°$ (012)/[100] LAMGB; (i3) $\phi = 1.77°$ (012) LATwGB.

Disconnection network structures of eight LAMTGB in zones α, β and γ are given in Figure 5, where Figures 5a and 5b show the LAMTGBs with hexagonal disconnection networks that are similar to (111) shuffle LATwGB in Figure 4g3. LAMTGBs in Figures 5c, 5e and 5f are simple disconnection networks, while the other two in Figures 5d and 5g include additional stacking faults. All LAMTGBs, except for the one in Figure 5b, are incoherent disconnection networks, which means that their boundary plane has deviated from the symmetric plane in the direction that is vertical to the ⅙<211> screw dislocation line. Comparing Figures 5e and 5f located on the *TTR* = 6 line in Figure 5h, it indicates that LAMTGB shares a common feature with LAMGB: disconnection network topology does not vary with the parameter *TTR*. Using such a conclusion and the results of Wan and Tang [66], we can approximate the LAMGB and LAMTGB portions in the Rodriguez-Frank space as a spherical pyramid with radius $R$ (determined by the average dislocation core radius) in Figure 5i to estimate their populations.

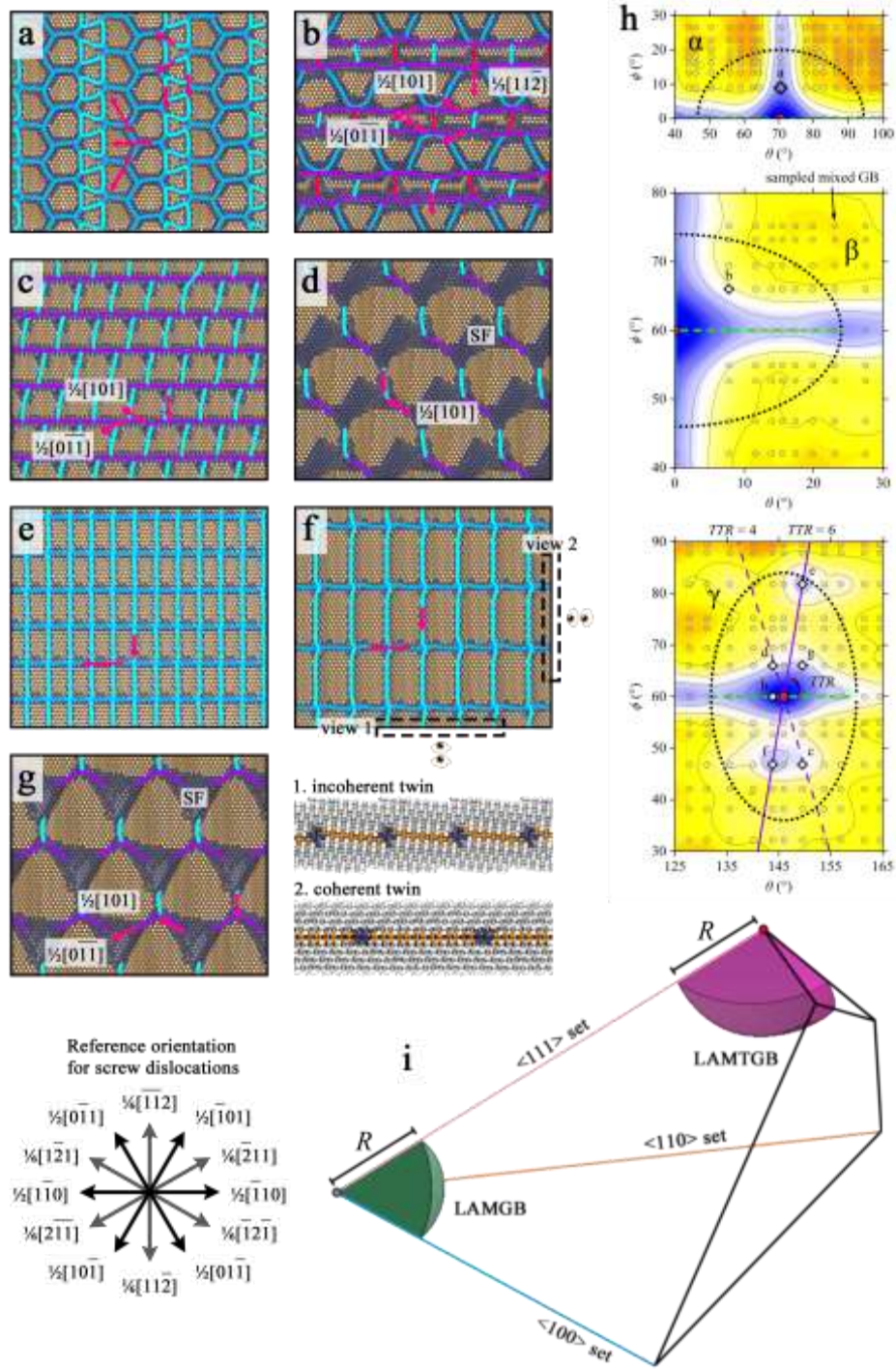

**Figure 5.** Disconnection network structures in the LAMTGB zones α, β and γ of Figure 3. (a)-(g) 2D planar view of the LAMTGB structures, where (a), (c), (d), (e), (f) and (g) are incoherent type, and (b) is coherent type. An example of the incoherent disconnection network is attached with (f); (h) Positions of (a)-(g) in the mixed character space. (i) Portions of LAMGB and LAMTGB in the Rodriguez-Frank space. Components of LAMGB (a) to (g) are not given because of the following paradox: twin GB may be the tilt or the twist component, depending on the selection of the decomposition axis. Explanations of symbols follow the same in Figure 4.

Interestingly, in all cases, despite notable restructuring, the mixed GBs still possess notable structural characteristics from their components. This is even true in the case of the twin GB that could accommodate complex disconnection structures, and the role of the twin GB is just like the perfect crystal in the mixed character space, as illustrated in Figure 5i. The authors are not only optimistic about these structural features of LAMGBs and LAMTGBs due to their promising usage in self-assembled nanostructures but also expect unique insights from their migration behaviors.

*3.2.2. Special HAMGB structures*

Structures of four HAMGBs with typical dislocation stress fields but without identifiable dislocation core structures are plotted in Figure 6. Figure 6a1 shows a near Σ5 HAMGB structure generated by introducing a 4.77° twist angle on the Σ5 "kite" STGB (Supplementary Figure S1a) along the boundary plane. Figure 6a2 shows a squared shadow pattern, which shape is similar to the dislocation network of a $\phi = 4.77°$ (001) LATwGB in Figure 6a4. HAMGB stress fields in Figure 6a3 are also similar to the LATwGB stress fields in Figure 6a5. Figure 6b1 shows a near Σ15 HAMGB structure generated by introducing a 4.13° twist angle on the Σ15 STGB, while Figures 6b2 and 6b4 are the upper views of the HAMGB and its twist component $\phi = 4.13°$ (111) LATwGB, respectively. Different but considerable stress fields are shown in Figure 6b3, compared with the counterpart in Figure 6b5. Such differences are explained by the superposition and mutual perturbation of the stress fields from the HAMGB components.

To verify such an explanation, Figures 6c1 and 6d1 show two HAMGBs, both of which share the same tilt components ($\theta = 4.77°$ (001)/[100] LASTGB) shown in Figure 6c5. Their stress fields in Figures 6c2 and 6d2 show similar characteristics with the dislocation stress fields in Figure 6c6, although both structures are extended without identifiable dislocations or SUs (i.e., SU is too big so the SU model is not applicable, similar to the prediction of the very early inter-crystalline amorphous cement theory [92]). It should be noted that the stress fields in Figure 6d2 are less evident than in Figure 6c2 in the dislocation characteristics, which is explained by comparing the structural complexity (and stress fields) of their twist components shown in Figures 6c3 and 6d3 (Figures 6c4 and 6d4).

By comparing two cases (i) STGB accommodates a low angle twist component and (ii) TwGB accommodates a low angle tilt component, it is found that the stress fields of the mixed GB with a low angle component contain stress characteristics from that low angle component, and such stress characteristics are more evident in simpler SUs. It also suggests that the contribution of elastic strain should be considered for the energies of these extended GB structures, rather than focusing on lattice distortion energy conventionally. Thus, while the HAMGBs do not contain as easily identifiable structural characteristics of their components, like the LAMGBs and LAMTGBs, the stress characteristics do show the contributions from the components.

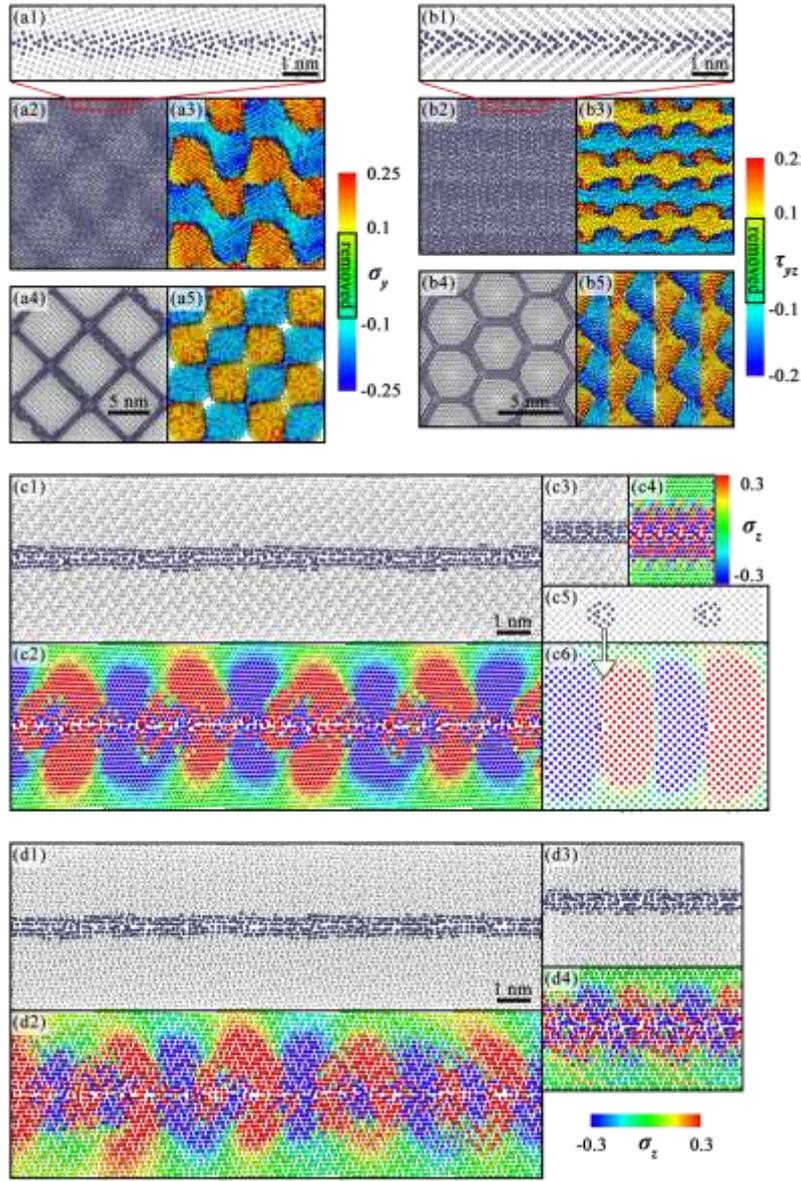

**Figure 6.** Atomic structures and stress fields of special HAMGBs in Figure 3. (a1) Side view of atomic structures of Σ26001 (288 12 577) HAMGB, which is formed by introducing 4.77° twist angle on $\theta$ = 53.13° Σ5 (102) STGB; (a2) Upper view of atomic structures of Σ26001 (288 12 577) HAMGB; (a3) Upper view of distribution of stress component $\sigma_y$ on (a2); (a4) Upper view of atomic structures of $\phi$ = 4.77° (001) LATwGB; (a5) Upper view of distribution of stress component $\sigma_y$ on (a4); (b1) Side view of atomic structures of Σ328363 (507 5 267) HAMGB, which is formed by introducing 4.13° twist angle on $\theta$ = 78.46° Σ15 (201) STGB; (b2) Upper view of atomic structures of the Σ328363 (507 5 267) HAMGB; (b3) Upper view of distribution of stress component $\tau_{yz}$ on (b2); (b4) Upper view of atomic structures of $\phi$ = 4.13° (111) LATwGB; (b5) Upper view of distribution of stress component $\tau_{yz}$ on (b4); (c1) Side view of atomic structures of Σ3605 (2 1 60) HAMGB, which is formed by introducing 4.77° tilt angle on $\phi$ = 36.87° Σ5 (001) TwGB; (c2) Side view of distribution of stress component $\sigma_z$ on (c1); (c3) Side view of atomic structures of $\phi$ = 53.13° Σ5 (012) TwGB; (c4) Side view of distribution of stress component $\sigma_z$ on (c3); (c5) Side view of atomic structures of $\theta$ = 4.77° (001)/[100] LASTGB; (c6) Side view of distribution of stress component $\sigma_z$ on (c5); (d1) Side view of atomic structures of a Σ485141 (25 10 696) HAMGB, which is formed by introducing 4.77° tilt angle on $\phi$ = 43.60° Σ29 (025) TwGB; (d2) Side view of distribution of stress component $\sigma_z$ on (d1); (d3) Side view of atomic structures of $\phi$ = 43.60° Σ29 (025) TwGB; (d4) Side view of distribution of stress component $\sigma_z$ on (d3).

### 3.2.3. Comparison with experiments

Figure 7 compares experimental observations and atomistic simulations for four silicon GBs, and the simulated GB characters are marked with stars in Figure 3. Surprisingly, the simulation shows the capability to accurately reproduce both dislocation and extended silicon GB structures at the atomic level. For example, experiment versus simulation is 1:1 in the length scale for GBs in Figures 7a and 7d. The simulation reproduces the complex dislocation network topology and even captures the meta-stable dislocation network that is partially shown in Figure 7a1. For the experimentally observed LAMGBs in Figures 7b1 and 7c1, direct comparison is inappropriate because their sizes usually make the computation resources unaffordable. The comparisons with simulation are addressed on the basis of topology since dislocation network topology does not vary with *TTR*. This allows us to simulate topological equivalent structures in a length ratio like 1:2.66 or 1:7. The presented simulations in Figures 7b2 and 7c2 agree well with the experimental counterparts, while only ignoring a slight dissociation at the triple junctions of (111) shuffle LATwGB at finite temperatures. The dissociation causes the difference and mutual conversion between (111) shuffle and glide LATwGB in diamond and FCC lattices [45, 93, 94]. For the SU GB in Figure 7d1, the simulation in Figure 7d2 gives an excellent agreement, the error of each atom position is sub-angstrom level. In summary, the comparisons between experiments and simulations show remarkable similarity. This provides support for the methodology and expectation that mixed GBs should exhibit characters of their components.

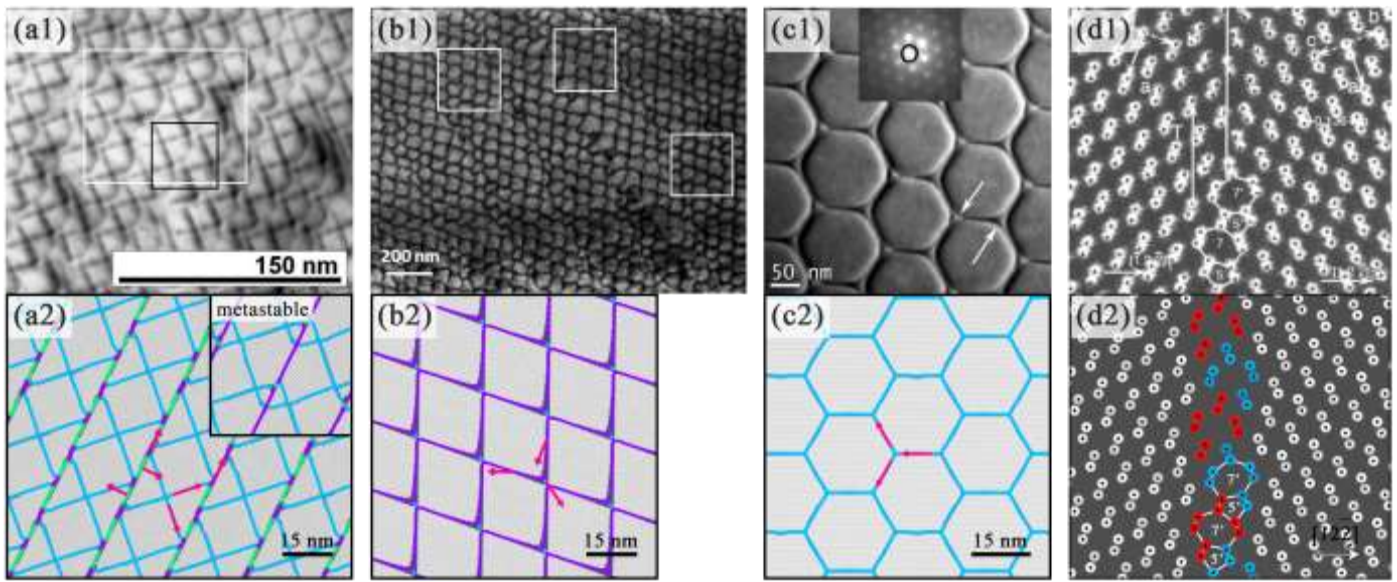

**Figure 7.** Comparisons between experimental observations and atomistic simulations. (a1) Experimental observation of a (001) LAMGB from Wilhelm et al. [81]; (a2) 1:1 simulation reproduction of the stable (001) LAMGB structures, the subfigure indicates a meta-stable (001) LAMGB structures; (b1) Experimental observation of a (011) LAMGB from Reiche et al. [95]; (b2) 1:2.66 simulation reproduction of the (011) LAMGB structures; (c1) Experimental observation of a (111) LATwGB from Neily et al. [76]. The hexagonal dislocation network partially dissociates to triangular hybrids of dislocation and stacking fault at the triple junctions; (c2) 1:7 simulation reproduction of the (111) shuffle LATwGB structures, which energy is slightly lower than (111) glide LATwGB structures; (d1) Experimental observation of a (011) HASTGB from Bonnet, et al. [96]; (d2) 1:1 simulation reproduction of the (011) HASTGB structures. The error of each atom position is less than 0.01 nm; Noting that experiments only determined the boundary planes, and thus the GB descriptions here ignore the misorientation axis (e.g., (001)/[100] LAMGB turns to (001) LAMGB).

### 3.3. An extended Read-Shockley model

*3.3.1. Theorization*

As discussed in sections 3.1 and 3.2, the trends of LAMGB energy surface could be described by the Read-Shockley relationship due to the dislocation characteristics. Structural analysis of LAMTGB and special HAMGB indicates that their energies are expected to have similar (predictable) trends once twin/SU GBs are set as the reference lattice. Therefore, the Read-Shockley relationship could be extended for LAMTGB and special HAMGB, from which a generalized framework for predicting mixed GB energies is proposed.

Figure 8a shows the assumed energy trends in a mixed character space. Different types of Read-Shockley relationships are assumed in LAMGB, LAMTGB and special HAMGB zones where the GB energies are considered predictable. The energy trends of the general HAMGB zone are not considered for prediction because the disappearance of dislocation core structures makes the energy trend rugged and invalidates the theory underlying Read-Shockley relationships.

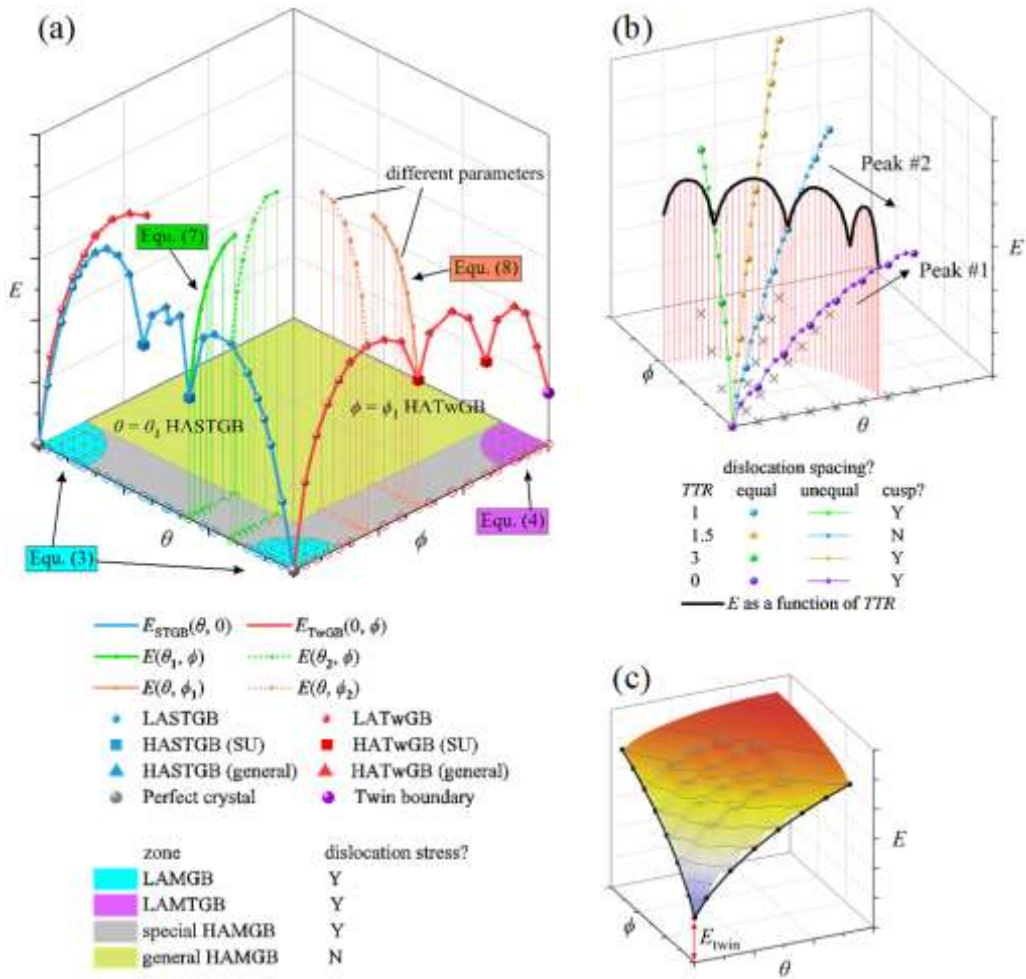

**Figure 8.** Conceptualization and illustration of the extended Read-Shockley model. (a) Assumed energy trends in an assumed mixed character space, where the energies of LAMGB, LAMTGB and special HAMGB zones are predictable; (b) Assumed GB energy trends in an assumed LAMGB zone as a 2D extension of the classical Read-Shockley relationship. The cusps depend on *TTR*; (c) Assumed 2D energy surface in an assumed LAMTGB zone; Note that all data in Figure 8 is not real due to the illustration purpose.

Then, we consider the factor that affects the Read-Shockley relationships of LAMGBs. For a given LAMGB, we need to realize that it can be indexed by two manners (i) tilt angle $\theta$ and twist angle $\phi$; (ii) tilt-twist ratio *TTR* and a combined angle *A* that measures the total contribution of both θ and ϕ following:
$$A = \sqrt{\theta^2 + \varphi^2} \tag{9}$$
*A* approximately equals to misorientation angle when θ → 0 and ϕ → 0. Sometimes *TTR* and *A* are useful to characterize a LAMGB because *TTR* determines the dislocation network topology and *A* scales the dislocation network size. With these physical facts, we can elucidate that the Read-Shockley relationships of (and fitting parameters within) a set of LAMGBs are associated with their *TTR*. However, such a conclusion does not explain the complex ground truth of LAMGB zones, especially for the Read-Shockley cusps marked in Figure 3. Therefore, the assumed LAMGB energy trends along several specific trace lines (*TTR* = constants) in a mixed character space are plotted in Figure 8b for explanation. Although the LAMGBs on a given trace line share the same dislocation network topology, unequal dislocation spacing would appear in most of them (the medium is not continuous at the atomic level) and generate peaks (peak type #1 in Figure 8b as a function of *A*), as highlighted by Read and Shockley decades ago [48]. For example, for the *TTR* = 0 (purple) line in Figure 8b, cusp behavior is expected on some GBs marked by large purple markers. However, Read and Shockley did not indicate the role of *TTR*: **LAMGB energies (e.g., Read-Shockley cusps) are especially sensitive to *TTR*.** When some special *TTR* is not satisfied by the two LAMGB components, excess dislocations are generated and then generate peaks (peak type #2 in Figure 8b as a function of *TTR*, which is illustrated by the black line) between these special *TTR* (assuming *A* is fixed). On the contrary, two components are perfectly "mixed" without excess dislocations under the special *TTR*, such as 1 or 3, where significant cusp behavior is exhibited (An example is given in Supplementary Materials). The special *TTR* can be predicted from LASTGB and LATwGB structures and the reconstruction mechanisms, or directly using FBE in some cases. It should be noted that the excess energy (peak type #2) from *TTR* is always greater than the excess energy (peak type #1) from *A*. In other words, LAMGB energy suffers low and high peaks in the directions (i) variable *A* and fixed *TTR* and (ii) fixed *A* and variable *TTR*, respectively. These discussions also explain the complex landscapes in LAMTGB zones α, β and γ due to their similarity with LAMGB. Figure 8c shows an assumed LAMTGB energy surface. One must notice that the origin contains a twin GB energy, which may be ignored for silicon but not for other materials if significant. Finally, for the special HAMGB with dislocation stress, we only consider the situation that a low tilt/twist angle is introduced on a SU GB. In that case, the Read-Shockley relationship only needs to add an energy term for the SU GB as the "substrate".

Recent work by some of the authors has shown that the energies of LAMGBs are equal to the sum of the dislocation core and strain energies of their tilt and twist components minus the variations from structural reconstruction, which can also be divided in terms of dislocation core and strain energies [66]. This energy *E* for LAMGB can be analytically written as the function of θ and ϕ following:
$$E_{\text{LAMGB}}^{\text{Total}}(\theta,\phi) = E_{\text{LASTGB}}^{\text{Total}}(\theta) + E_{\text{LATwGB}}^{\text{Total}}(\phi) + E_{\text{Reconst}}^{\text{Core}}(\theta,\phi) + E_{\text{Reconst}}^{\text{Strain}}(\theta,\phi) \tag{2}$$
Where superscript Total, Core and Strain denote the LAMGB total excess energy, dislocation core and strain energies, respectively. Subscript LAMGB, LASTGB and LATwGB denote the GB type. Subscript Reconst denotes the energy variations of the structural reconstruction. Read and Shockley have already given the analytical solution of the LASTGB and LATwGB energies known as the classical Read-Shockley relationship (θ − θln(θ) and ϕ − ϕln(ϕ)) for the first two terms of equation (2), but the detailed expressions of the last two terms are still unclear. A semi-empirical expression of the last two terms has been given by Wan and Tang [66] to revise the Read-Shockley relationship for LAMGB following:

$$E_{\text{LAMGB}}^{\text{Total}}(\theta,\phi) = \theta\left[E_{\text{LASTGB}}^{\text{Core}} - E_{\text{LASTGB}}^{\text{Strain}}\ln(\theta)\right] + \phi\left[E_{\text{LATwGB}}^{\text{Core}} - E_{\text{LATwGB}}^{\text{Strain}}\ln(\phi)\right] + \theta\phi\left[E_{\text{Reconst}}^{\text{Core}} - E_{\text{Reconst}}^{\text{Strain}}\ln(\theta\phi)\right] \quad (3)$$

Where the last two terms $E_{\text{Reconst}}^{\text{Core}} + E_{\text{Reconst}}^{\text{Strain}}$ that follow the Read-Shockley formalism $\theta\phi - \theta\phi\ln(\theta\phi)$ are considered to have a good balance of both fitting effects and clarity [66] Additional explanation of the rationality of this form is given in Supplementary Materials, but in short, it agrees with the prediction of Sutton and Balluffi that any LAMGB structure-energy relationships should follow or be similar to the Read-Shockley formalism [97]. The compatibility of equation (3) is robust. For example, by replacing $\theta$ and $\phi$ with $\sin(\theta)$ and $\sin(\phi)$, Equation (3) becomes Wolf's empirical equation of the Read-Shockley relationship [98]. An additional energy $E_i$ could be added to describe the dislocation interaction energy for the LATwGB, as Schwartz et al. [99] did. Here, these earlier revisions [100, 101] of the Read-Shockley relationship are not considered for simplicity (e.g., applying these revisions only increases limited performance but complicated function itself). Since LAMTGB structures are based on twin GB and their reconstructions are similar to LAMGB, an additional twin GB energy $E_{\text{Twin}}$ is appended to equation (3) to capture the LAMTGB energies following:

$$E_{\text{LAMTGB}}^{\text{Total}}(\theta_{\text{Twin}},\phi_{\text{Twin}}) = E_{\text{LAMGB}}^{\text{Total}}(\theta_{\text{Twin}},\phi_{\text{Twin}}) + E_{\text{Twin}} \quad (4)$$

Where $\theta_{\text{Twin}}$ and $\phi_{\text{Twin}}$ are the tilt and twist angles defined on the reference twin GB. Since some special HAMGBs (in Figure 6) also have dislocation stress characteristics (strain energy), we can extend equation (4) for the special HAMGBs. Considering the cases that a low twist angle $\phi$ is introduced on a $\theta = \theta_1$ HASTGB which energy is $E_{\text{HASTGB}}$) and a low tilt angle $\theta$ is introduced on a $\phi = \phi_1$ HATwGB which energy is $E_{\text{HATwGB}}$, then equation (4) varies to

$$E_{\text{HAMGB}}^{\text{Total}}(\phi) = E_{\text{LAMGB}}^{\text{Total}}(\theta_1,\phi) = E_{\text{HASTGB}} + \phi\left[E_{\text{LATwGB}}^{\text{Core}} - E_{\text{LATwGB}}^{\text{Strain}}\ln(\phi)\right] + \theta_1\phi\left[E_{\text{Reconst}}^{\text{Core}} - E_{\text{Reconst}}^{\text{Strain}}\ln(\theta_1\phi)\right] \quad (5)$$

and

$$E_{\text{HAMGB}}^{\text{Total}}(\theta) = E_{\text{LAMGB}}^{\text{Total}}(\theta,\phi_1) = \theta\left[E_{\text{LASTGB}}^{\text{Core}} - E_{\text{LASTGB}}^{\text{Strain}}\ln(\theta)\right] + E_{\text{HATwGB}} + \theta\phi_1\left[E_{\text{Reconst}}^{\text{Core}} - E_{\text{Reconst}}^{\text{Strain}}\ln(\theta\phi_1)\right] \quad (6)$$

respectively. By cancelling the known terms, the straightforward forms of equations (5) and (6) are written as

$$E_{\text{HAMGB}}^{\text{Total}}(\phi) = E(\theta_1,0) + \phi\left[E^{\text{Core}} - E^{\text{Strain}}\ln(\phi)\right] \quad (7)$$

and

$$E_{\text{HAMGB}}^{\text{Total}}(\theta) = \theta\left[E^{\text{Core}} - E^{\text{Strain}}\ln(\theta)\right] + E(0,\phi_1) \quad (8)$$

respectively. Where $E^{\text{Core}}$ and $E^{\text{Strain}}$ are fitting terms in the classical Read-Shockley relationship [46], and equations (7) and (8) finally degenerate to the classical Read-Shockley formalism. Noting that (i) the fitting terms vary with the mixed character space ($E^{\text{Core}} = f_1(\theta,\phi)$ and $E^{\text{Strain}} = f_2(\theta,\phi)$), thus predicting complex energy landscape of HAMGB zone may be difficult for equations with fixed parameters (an example that predicts a smooth energy surface is given by Wolf [67]). In summary, the energies of LAMGBs, LAMTGBs and special HAMGBs (all have dislocation stress) can be described in the framework of the extended Read-Shockley model.

*3.3.2. Performance*

The performance of the extended Read-Shockley model is examined on the presented GB dataset. Figures 9a1 and 9a2 plot the energy predictions for LAMGBs, LAMTGBs and special HAMGBs by fitting the model equations (3), (4) and (7), respectively. It is shown that the rising energy trends with both angles in LAMGB and LAMTGB zones are captured by equations (3) and (4), but the deep trench in LAMGB zone III is ignored since both equations only predict the generalized average of LAMGB and LAMTGB energy surfaces. Such results are consistent with the Read-Shockley relationship because Read and Shockley had given a good example [48]: clarified the ground truth that the LAGB energy curve is rugged as a supplementary to their classical relationship, which predicts LAGB energy as a smooth function of misorientation angle. Figure 9a1 also shows the energy trends along a specific slice of the HAMGB zone that contains a set of special HAMGBs formed by introducing low twist angles on the Σ5 STGB with well-known "kite" SU. The trend described by equation (7) is confirmed on these special HAMGB energies, which not only agrees with the observation of dislocation stress characteristics in Figure 6 but also implies that the other special HAMGBs (near low energy STGB or TwGB) are likely to follow the same Read-Shockley trend. Interestingly, Wolf had assumed similar smooth energy trends decades ago without examining the massive GB dataset like this [67].

Figure 9b is the parity plot comparing predicted energies versus computed energies for all examined GBs, including LAMGB zones I to V and LAMTGB zones α and β (Fitting parameters are given in Supplementary Tables S9, S10 and S11). Except for the Read-Shockley cusps, the model equations generally capture the trends of energy surfaces of the seven zones. The overall RMSE is 46.51 mJ/m$^2$ for 569 examined GBs in total, meanwhile, the minimum and maximum RMSEs of the seven zones that input for prediction are 24.75 mJ/m$^2$ (zone I) and 62.49 mJ/m$^2$ (zone II), respectively.

The extended model is designed to predict energies based on the dislocation characteristics of GBs. Although it has a nice fit for LAMGB, LAMTGB and special HAMGB with dislocation stress, it is not recommended for the general HAMGB zone due to the disappearance of dislocation characteristics. Given that the HAMGB has both high tilt and twist angles, the dislocation cores will overlap and result in considerable reconstruction where the dislocation model of grain boundaries would not be expected to apply. Nonetheless, it is shown that the Read-Shockley relationship does provide a reasonable prediction of high angle GB energies [42, 43, 66, 71, 97, 98], so we investigate that possibility here.

Figure 9c shows silicon GB energies calculated in this work as a function of misorientation angle, accompanied by the three types of Read-Shockley relationship (classical, Wolf's equation and equation (3)). The fitting of equation (3) is a group of orange points because it outputs an energy surface as a function of $\theta$ and $\phi$ in Figure 9d that describes the generalized average of the four energy surfaces in Figure 3. Meanwhile, the fitting process is to convert the energy surface in Figure 9d as a function of the misorientation angle by extracting *A* as the misorientation angle using equation (9) and ignoring *TTR*. In other words, it looks like that multiple slices are taken on the energy surface. In so doing, equation (3) would predict a cloud-like distribution showing the possible range of GB energy at any misorientation angles, as illustrated in Figure 9d. It is clear that this cloud-like distribution doesn't capture the full range of energies in all positions, but equation (3) exhibits natural advantages to capture the general trends formed by thousands of individual silicon GB energies in Figure 9c, compared with the other two Read-Shockley relationships that consist of a single curve.

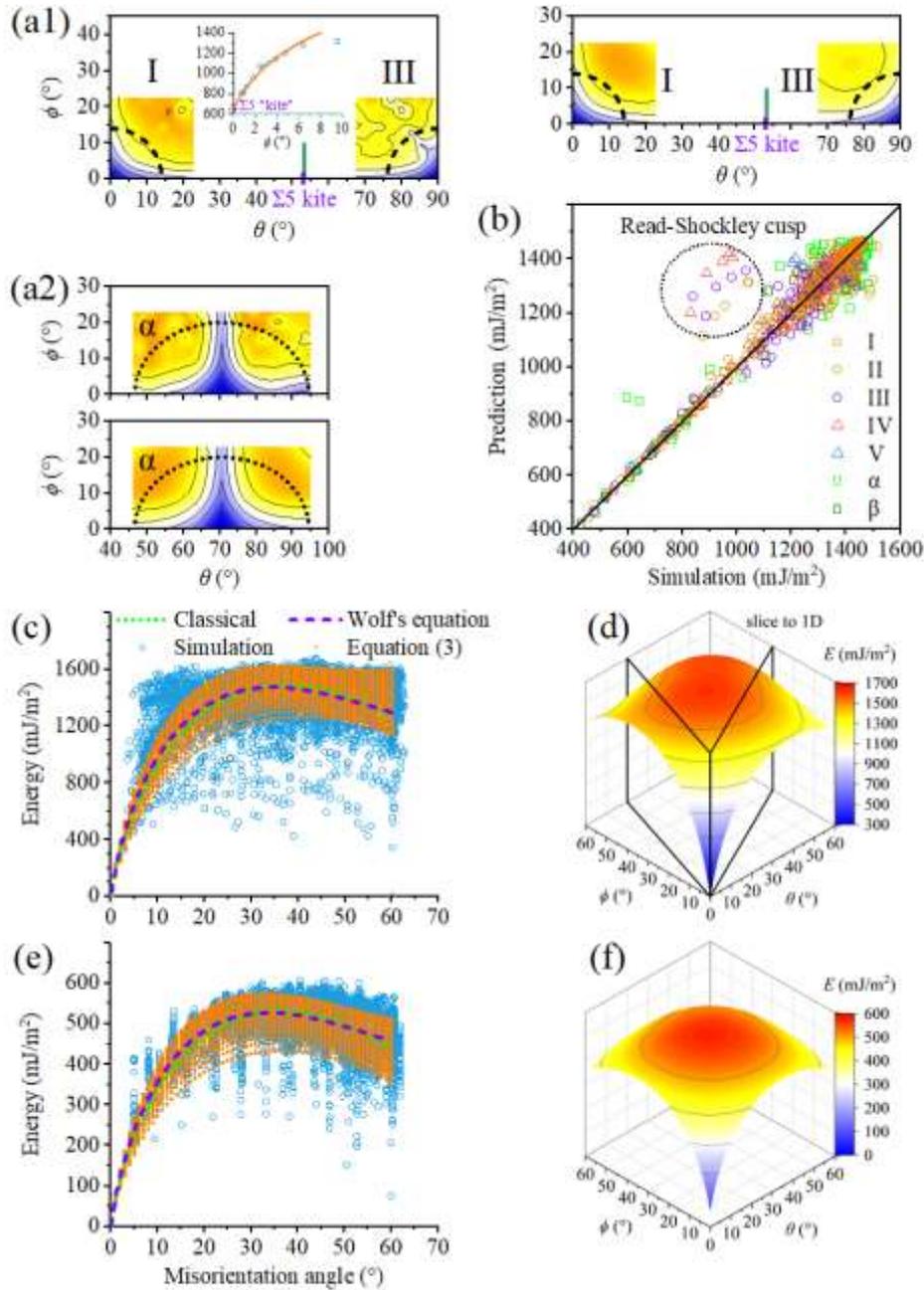

**Figure 9.** (a1) Simulated and predicted GB energies of LAMGB zones I and III, and a set of special HAMGBs near the well-known Σ5 "kite" STGB. Left is simulation and right is prediction; (a2) Simulated and predicted GB energies of LAMTGB zone α. Upper is simulation and lower is prediction; (b) Parity plot comparing predicted energies versus computed energies in LAMGB zones I to V and LAMTGB zones α and β. The Read-Shockley cusps where our model outputs bad predictions are marked with the dashed circle; (c) Silicon GB energies presented in this work as a function of the misorientation angle. Classical and Wolf's Read-Shockley relationships are fitted to data as the green and purple dashed lines, respectively. Equation (3) is fitted to data as the orange dots by converting the predicted energy surface in (d) to simplified descriptions; (d) Predicted energy surface for the silicon GB energies as a generalized average of the four energy surfaces in Figure 3; (e) Aluminum GB energies (from the work of Homer et al. [65]) as a function of the misorientation angle, accompanied with the three Read-Shockley relationships; (f) Predicted energy surface for the aluminum GB energies that used to fit data in (e).

The transferability of the extended model is tested on a recently published aluminum GB dataset [65]. The three types of Read-Shockley relationship are fitted in Figure 9e to compare their fitting effects and further confirm whether equation (3) is valid for FCC aluminum GBs. We can see that equation (3) almost perfectly captures the general trends of aluminum GB energy. The success in predicting both silicon and aluminum GB energy trends could be expanded to a wide range of FCC and diamond lattice materials through the key fact that GB energy is scaling with the elastic modulus in the materials with the same lattice [99]. One could also consider examining the validity of the BRK energy function [101, 102] over LAMGB zones and test the function performance on describing the complex landscapes of HAMGB zones.

## 4. Conclusions

Structures and energies of 7040 silicon GBs with tilt, twist and mixed tilt-twist characters are studied in this work to confirm the correlations between mixed GBs and their decomposed GB components. First of all, a qualitative and empirical strategy for predicting mixed GB properties from the tilt and twist components is given in Figure 11, which is helpful as a summary of the discussions.

| Tilt | | high angle | | low angle | | Relative energy | Structure type | Dislocation stress |
|---|---|---|---|---|---|---|---|---|
| Twist | | Ext. Struct. | SU | Dis. Struct. | | High | Ext. Struct. | N |
| high angle | Ext. Struct. | | | | | | SU + Dis. Struct. | Y |
| | SU | | | | | | SU | N |
| low angle | Dis. Struct. | | | | | Low | Dis. Struct. | Y |

**Figure 10.** Qualitative and empirical prediction strategy of mixed GB structures, energies and stress fields through the tilt and twist components. Ext. Struct.: extended structure; SU: structural unit; Dis. Struct.: dislocation structure. Extended structure mixed with any structures will result in a relatively high energy, and SU mixed with SU usually yields a SU. LAMGBs and LAMTGB are dislocation structures with the relatively lowest energy.

There are several other conclusions worth noting:

- Mixed GB structures contain dislocation/disconnection characteristics when a component is low angle type, and the proportions of characteristics are related to the *TTR* parameter. LAMGB structures are formed by the dissociation, motion and reaction mechanisms between the superposed (dislocation and stacking fault) characteristics of the two components. Such mechanisms reconstruct the structures and naturally reduce the energies. Coherent and incoherent LAMTGB structures follow the same mechanisms on their disconnection networks. By introducing a low angle tilt or twist component, typical dislocation stress fields are observed near some SU GBs, such as Σ5 "kite" GB, implying the wide existence of unidentifiable disconnections in HAMGBs with extended GB structures.
- Mixed GB energies can be predicted from the energies of their tilt and twist components and energetic variations from the structural reconstruction mechanisms. Meanwhile, they are also sensitive to the *TTR* parameter due to its effects on the GB structure. Combining these facts with the original conception of Read and Shockley, an extended Read-Shockley model is proposed for mostly mixed GBs. Compared with the classical Read-Shockley relationship, this model can predict the energy surfaces of LAMGBs and LAMTGBs or extend to any HAMGB with a low angle component. Most importantly, it predicts the general trends of GB energy in a uniquely accurate manner. The model transferability is also examined and proven on a recently published aluminum GB dataset.

- Experimentally observed silicon LAMGB and STGB structures are reproduced by the simulation and thus validate the simulation and the derived conclusions. In comparison with several HRTEM images showing stable/metastable dislocation network structures of two LAMGB and atom arrangements of a STGB, our simulation shows remarkable agreement, which provides strong confidence for the reported silicon GB datasets. First-principles calculation on the energies of low $\Sigma$ GBs is in good scaling with the atomistic simulation, which not only connects both simulation methods but further reinforces the reliability of the studied energetic properties.

From these conclusions, a short understanding about this work and its prospects is given as the following:

In some cases, mixed GB is called "mixed" because it contains characteristics of its components, and thus some correlations are indeed easy to anticipate. However, we present various structural and energetic correlations of dislocation/disconnection-structured LAMGB and LAMTGB with different misorientation axes and boundary planes, and demonstrate the correlations even exist for general GBs to highlight the unexpected results. Rugged trends of the energy surface formed by thousands of individual GB energy enrich the earlier work of Wolf and extend the prediction of Read and Shockley decades ago. At the same time, the trends and the fact that disconnection connects GBs that are crystallographic-close jointly imply the unpredictable nature of HAGB energy. The contributions of this work could be quite fundamental because it paved the way for understanding complex GB characters from the basic structure and energy, which is essential for the subsequent works focusing on other properties, such as migration, segregation and strength. It can be foreseen that the migration behaviors of the reported mixed GBs would be extraordinary.

Beyond these findings on this GB dataset, more intriguing features could be explored further. Our work only addressed the mixed symmetric tilt and twist GBs. It is uncertain whether the mixed asymmetric tilt and twist GBs follow the same correlation as the symmetric counterparts. For that case, an additional asymmetric tilt angle should be introduced for characterization, which would greatly complicate the results. Due to the non-standard representation and non-Euclidean features of the mixed character space, FZ representation will certainly bring new unique insights into the low angle mixed asymmetric tilt and twist GBs in the future.

## Data availability

Numerical data are available upon reasonable request.

## Acknowledgements

W. Wan acknowledges insightful discussions with Prof. J.B. Yang from the Institute of Metal Research, Chinese Academy of Science. W. Wan and C.X. Tang thank Prof. W.N. Zou from Nanchang University for his support.

## Competing interests

The authors declare no competing interests.

# Fundings

C.X. Tang was supported by the National Natural Science Foundation of China (grant number: 11802112).

E.R. Homer was supported by the U.S. National Science Foundation (NSF) under Award #DMR-1817321.

W. Wan was supported by the Institute for Advanced Study of Nanchang University (nominal, not received).

# Author contributions

(i) W. Wan carried out this project, performed all simulations, visualized all figures, theorized the extended Read-Shockley model, wrote the original manuscript and revised it.

(ii) C.X. Tang supervised the project, contributed computation resources, maintained the data and revised the manuscript.

(iii) E.R. Homer provided the fundamental zone description of the dataset, discussed the data and revised the manuscript.